\RequirePackage{fix-cm}
\documentclass[twocolumn,epjc3]{svjour3}

\smartqed  
\usepackage{graphicx}
\usepackage{amssymb}
\usepackage[T1]{fontenc} 
\usepackage[utf8]{inputenc}
\usepackage[normalem]{ulem}
\usepackage{todonotes}
\usepackage{hyperref}

\newcommand{\mnu}{\ensuremath{\Sigma m_\nu}}

\newcommand{\lcdm}{$\Lambda$CDM}
\newcommand{\lcdme}{\ensuremath{\Lambda\mbox{CDM}}}
\newcommand{\mcm}[1]{\ensuremath{\mathcal{M}_{#1}}}
\newcommand{\eqref}[1]{(\ref{#1})}

\journalname{Eur. Phys. J. C}

\begin{document}

\title{Constraining power of open likelihoods, made prior-independent}


\author{S.\ Gariazzo}


\institute{Instituto de F\'{\i}sica Corpuscular (CSIC-Universitat de Val\`{e}ncia)\\
Parc Cient\'{\i}fic UV, C/ Catedr\'atico Jos\'e Beltr\'an, 2, E-46980 Paterna (Valencia), Spain
              \email{gariazzo@ific.uv.es}
}

\date{Received: date / Accepted: date}

\maketitle

\begin{abstract}
One of the most criticized features of Bayesian statistics
is the fact that credible intervals,
especially when open likelihoods are involved,
may strongly depend on the prior shape and range.
Many analyses involving open likelihoods are affected by
the eternal dilemma of choosing between linear and logarithmic prior,
and in particular in the latter case
the situation is worsened by the dependence on the prior range under consideration.
In this letter,
we revive a simple method to obtain constraints
that depend neither on the prior shape nor range
and,
using the tools of Bayesian model comparison,
extend it to overcome the possible dependence of the bounds
on the choice of free parameters in the numerical analysis.
An application to the case of cosmological bounds on the sum of the neutrino masses
is discussed as an example.
\end{abstract}

\section{Introduction}
In several cases,
physics experiments try to measure unknown quantities:
the mass of some particle,
a new coupling constant,
the scale of new physics.
Most of the times, the absolute scale of such new quantities
is completely unknown, and the analyses of experimental data require
to scan a very wide range of values for the parameter under consideration,
to finally end up with a lower or upper bound
when data are compatible with the null hypothesis.

In the context of Bayesian analysis,
performing this kind of analysis implies a profound discussion on the choice of the considered priors,
which may be logarithmic when many orders of magnitude are involved.
A robust analysis usually shows what happens when more than one type of prior is considered,
but the calculation of credible intervals always require also a precise definition of the prior range.
Especially in the case of logarithmic priors,
a choice of the range can be difficult even when physical boundaries
(e.g.\ a mass or coupling must be positive) exist,
with the consequence that the selected allowed range for the parameter
can influence the available prior volume and as a consequence the bound itself.

Let us consider for example the case of neutrino masses and their cosmological constraints.
Current data are sensitive basically only on the sum of the neutrino masses
and not on the single mass eigenstates
(see e.g.\ \cite{Lattanzi:2017ubx,deSalas:2018bym}).
There are therefore good reasons to describe the physics by means of \mnu\
and to consider a linear prior on it, as the parameter range is limited
from below by oscillation experiments \cite{Capozzi:2018ubv,deSalas:2017kay,Esteban:2018azc}
and from above by KATRIN \cite{Aker:2019uuj}.
Even given these considerations, however,
one can decide to perform the analysis considering a lower limit $\mnu>0$ \cite{Aghanim:2018eyx},
instead of enforcing the oscillation-driven one,
$\mnu\gtrsim60\,(100)$~meV (respectively for normal and inverted ordering of the neutrino masses,
see e.g.\ \cite{Wang:2017htc,Wang:2016tsz}):
the obtained \emph{upper} bounds will differ in the various cases.

In order to overcome these problems,
in this letter we revisit a simple way \cite{Astone:1999wp,DAgostini:2000edp,DAgostini:2003}
to use Bayesian model comparison techniques
to obtain prior-independent constraints,
which can be useful for an easier comparison
of the constraining power of various experimental results,
not only in the context of cosmology, but in all Bayesian analyses in general.
Furthermore, we extend the already known method to address the problems related
to the possible existence of degeneracies with multiple free parameters
and the choice of the considered parameterizations
when performing the numerical analyses.

\section{Prior-free Bayesian constraints}
The foundation of Bayesian statistics is represented by the Bayes theorem:
\begin{equation}\label{eq:bayestheorem}
p(\theta|d,\mcm{i})
=
\frac{\pi(\theta|\mcm{i}) \mathcal{L}_{\mcm{i}}(\theta)}{Z_i}\,,
\end{equation}
where
$\pi(\theta|\mcm{i})$ and $p(\theta|d,\mcm{i})$ are the prior and posterior probabilities
for the parameters $\theta$ given a model \mcm{i},
$\mathcal{L}_{\mcm{i}}(\theta)$ is the likelihood function, depending on the parameters $\theta$,
given the data $d$ and the model \mcm{i},
and
\begin{equation}\label{eq:bayesianevidence}
Z_i
=
\int_{\Omega_\theta}
d\theta\,\pi(\theta|\mcm{i})\,\mathcal{L}_{\mcm{i}}(\theta)
\end{equation}
is the Bayesian evidence of \mcm{i}~\cite{Trotta:2008qt},
the integral of prior times likelihood
over the entire parameter space $\Omega_\theta$.

While the Bayes theorem indicates how to obtain the posterior probability
as a function of all the model parameters $\theta$,
when presenting results we are typically interested in the marginalized posterior probability
as a function of one parameter (or two), which we can generally indicate with $x$.
The marginalization is performed over the remaining parameters, which we can indicate with $\psi$:
\begin{equation}\label{eq:marginalizedposterior}
p(x|d,\mcm{i})
=
\int_{\Omega_\psi}d\psi\, p(x,\psi|\mcm{i},d)\,.
\end{equation}
Let us now assume that the prior is separable and we can write
$\pi(\theta|\mcm{i})=\pi(x|\mcm{i})\cdot\pi(\psi|\mcm{i})$.
Under such hypothesis, Eq.~\eqref{eq:marginalizedposterior} can be written as:
\begin{equation}\label{eq:marginalizedposterior_explicit}
p(x|d,\mcm{i})
=
\frac{\pi(x|\mcm{i})}{Z_i}
\int_{\Omega_\psi}d\psi\, \pi(\psi|\mcm{i})\mathcal{L}_{\mcm{i}}(x,\psi)\,.
\end{equation}

Let us consider the marginalized posterior as written in Eq.~\eqref{eq:marginalizedposterior_explicit}.
The prior dependence is only present explicitly outside the integral,
and therefore we can obtain a prior-independent quantity %
\footnote{
This is not exactly true, in the sense that the prior also enters the calculation of the Bayesian evidence,
see Eq.~\eqref{eq:bayesianevidence}.
The shape of the posterior, in any case, is not affected by such contribution,
that only enters as a normalization constant.
}
just dividing the posterior by the prior.
The right-hand side of Eq.~\eqref{eq:marginalizedposterior_explicit}, however,
has an explicit dependence on the value of $x$ through the likelihood that appears in the integral.
We can note that such integral
resembles the definition of the Bayesian evidence in Eq.~\eqref{eq:bayesianevidence},
not anymore for model $\mcm{i}$,
but for a sub-case of $\mcm{i}$ which contains $x$ as a fixed parameter.
Let us label this model with $\mcm{i}^{x}$ and define its Bayesian evidence:
\begin{equation}\label{eq:bayesianevidence_x}
Z_i^{x}
\equiv
\int_{\Omega_\psi}d\psi\, \pi(\psi|\mcm{i})\mathcal{L}_{\mcm{i}}(x,\psi)\,,
\end{equation}
which is independent of the prior $\pi(x)$, but still depends on the parameter value $x$,
now fixed.
Note that Eq.~\eqref{eq:marginalizedposterior_explicit}
can be rewritten in the following form:
\begin{equation}\label{eq:Zi_prior}
Z_i
=
\frac{\pi(x|\mcm{i})}{p(x|d,\mcm{i})}
Z_i^x\,.
\end{equation}

Now, let us consider two models $\mcm{i}^{x_1}$ and $\mcm{i}^{x_2}$.
Since $Z_i$ is independent of $x$, we can use Eq.~\eqref{eq:Zi_prior} to obtain
\begin{equation}\label{eq:eq_x1_x2}
\frac{\pi(x_1|\mcm{i})}{p(x_1|d,\mcm{i})}
Z_i^{x_1}
=
\frac{\pi(x_2|\mcm{i})}{p(x_2|d,\mcm{i})}
Z_i^{x_2}
\,,
\end{equation}
which can be rewritten as 
\begin{equation}\label{eq:Zix_ratio}
\frac{Z_i^{x_1}}{Z_i^{x_2}}
=
\frac{p(x_1|d,\mcm{i})/\pi(x_1|\mcm{i})}{p(x_2|d,\mcm{i})/\pi(x_2|\mcm{i})}
\,.
\end{equation}
The left hand side of this equation is a ratio of the Bayesian evidences
of the models $\mcm{i}^{x_1}$ and $\mcm{i}^{x_2}$,
therefore it is a Bayes factor.
For reasons that will be clear later,
let us rename $x_1\rightarrow x$ and $x_2\rightarrow x_0$ and
define this ratio as $\mathcal{R}(x,x_0|d)$,
which was named ``relative belief updating ratio'' or ``shape distortion function'' in the past
\cite{Astone:1999wp,DAgostini:2000edp,DAgostini:2003}:
\begin{equation}\label{eq:R_definition}
\mathcal{R}(x,x_0|d)
\equiv
\frac
{Z_i^{x}}
{Z_i^{x_0}}
=
\frac{
p(x|d,\mcm{i})/\pi(x,\mcm{i})
}{
p(x_0|d,\mcm{i})/\pi(x_0,\mcm{i})
}
\,.
\end{equation}
Although this function has been already employed in the past,
see e.g.~\cite{D'Agostini:1999wq,Eitel:1999gt,Abreu:1999aa,Breitweg:1999ssa},
its use has been somewhat faded into obscurity.
Here, we will revise its properties and discuss them in details.

Let us recall that $Z_i^x$ is independent of $\pi(x)$, see Eq.~\eqref{eq:bayesianevidence_x}:
this means that $\mathcal{R}(x,x_0|d)$ is also independent of $\pi(x)$.
This quantity therefore represents a prior-independent way
to compare some results concerning two values of some parameter $x$.
At the practical level, $\mathcal{R}$ is particularly useful when dealing with open likelihoods,
i.e.\ when data only constrain the value of some parameter from above or from below.
In such case, the likelihood becomes insensitive to the parameter variations below (or above) a certain threshold.
Let us consider for example the absolute scale of neutrino masses,
on which data (either cosmological or at laboratory experiments)
only put an upper limit:
the data are insensitive to the value of $x$ when $x$ goes towards 0, so
we can consider $x_0=0$ as a reference value.
Regardless of the prior, when $x$ is sufficiently close to $x_0$ the likelihoods in $x$ and $x_0$
are essentially the same
in all the points of the parameter space $\Omega_\psi$,
so $Z_i^{x}\simeq Z_i^{x_0}$ and $\mathcal{R}(x,x_0)\rightarrow1$.
In the same way, when $x$ is sufficiently far from $x_0$, the data penalize its value
($Z_i^{x}\ll Z_i^{x_0}$)
and we have $\mathcal{R}(x,x_0)\rightarrow0$.
In the middle,
the function $\mathcal{R}$ indicates how much $x$ is favored/disfavored with respect to $x_0$ in each point,
in the same way a Bayes factor indicates how much a model is preferred with respect to another one.

While $\mathcal{R}$ can define the general behavior of the posterior as a function of $x$,
any probabilistic limit one can compute will always depend on the prior shape \emph{and range},
which is an unavoidable ingredient of Bayesian statistic.
The description of the results through the $\mathcal{R}$ function, however,
allows to use the data to define a region above which the parameter values are disfavored,
regardless of the prior assumptions,
and also to guarantee an easier comparison of two experimental results.
A good standard could be to provide a (non-probabilistic) ``sensitivity bound'',
defined as the value of $x$ at which $\mathcal{R}$ drops below some level,
for example $|\ln\mathcal{R}|=1$, 3 or 5 in accordance to the Jeffreys' scale
(see e.g.\ \cite{deSalas:2018bym,Trotta:2008qt}).
Let us consider $x_0=0$ as above:
we could say, for example, that we consider as ``moderately (strongly) disfavored'' the region $x>x_s$
for which $\ln\mathcal{R}<s$, with $s=-3$ (or $-5$),
and then use the different values $x_s$ to compare the strengths
of different data combinations $d$ in constraining the parameter $x$.
This will not represent an upper bound at some given confidence level, since it is not a probabilistic bound,
but rather a hedge ``which separates the region in which we are, and where we
see nothing, from the the region we cannot see'' \cite{DAgostini:2000edp}.

From the computational point of view, it is not necessary to perform the integrals
in the definition of $Z_i^x$ in order to compute $\mathcal{R}$.
One can directly use the right hand side of Eq.~\eqref{eq:R_definition},
i.e.\ numerically compute $p(x|d,\mcm{i})$ with a specific prior assumption,
then divide by $\pi(x,\mcm{i})$ and normalize appropriately.
Notice also that,
once $\mathcal{R}$ is known,
anyone can obtain credible intervals with any prior of choice:
the posterior $p(x|d,\mcm{i})$ can easily be computed using Eq.~\eqref{eq:R_definition}
and normalizing to a total probability of 1 within the prior.

\begin{figure}[t]
\centering
\includegraphics[width=\columnwidth]{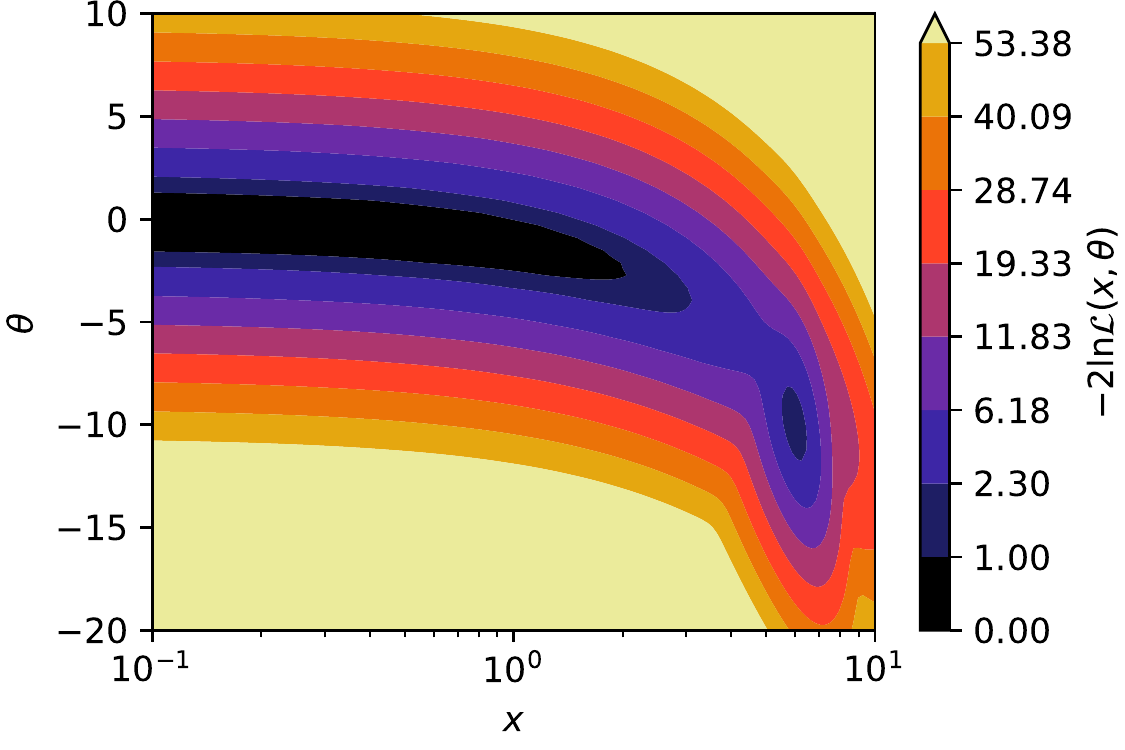}\\
\includegraphics[width=\columnwidth]{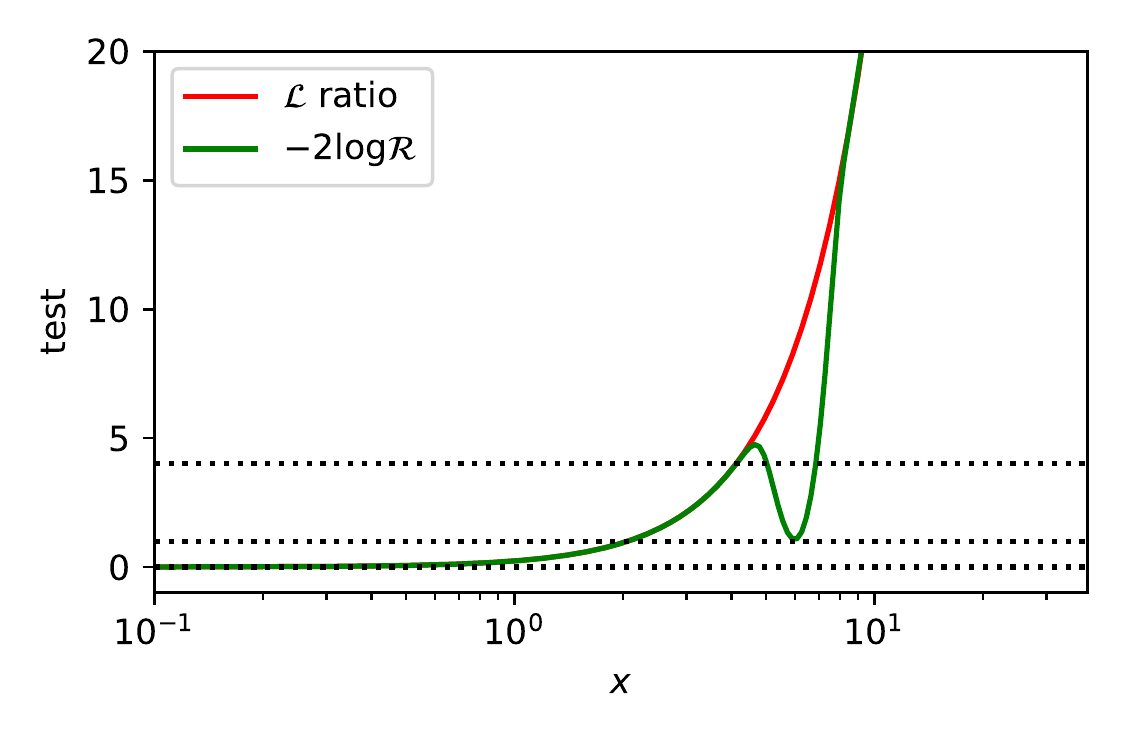}
\caption{\label{fig:R_vs_L_ex}
\textit{Upper panel:}
Dependence of the likelihood in Eq.~\eqref{eq:example_llh}
on the parameters $x$ and $\theta$.
\textit{Lower panel:}
Comparison of the results obtained with the likelihood ratio and the $\mathcal{R}$ methods
when the likelihood in Eq.~\eqref{eq:example_llh} is considered.
The horizontal lines show the levels $0$, $1$, $4$, respectively.
}
\end{figure}

Few final comments:
in most of the cases,
obtaining limits with the $\mathcal{R}$ function
is nearly equivalent to using a likelihood ratio test.
The difference is that, while the likelihood ratio test only takes into account
the likelihood values in the best-fit at fixed $x_0$ and $x$,
the $\mathcal{R}$ method weighs the information of the entire posterior distribution
and takes into account the mean likelihood over the prior $\Omega_\psi$.
This means that in cases with multiple posterior peaks or complex posterior distributions,
the limits obtained using the $\mathcal{R}$ function can be more conservative
than those obtained with the likelihood ratio test.
As an example, we provide in the lower panel of Fig.~\ref{fig:R_vs_L_ex} a comparison of the
likelihood ratio and of the $-2\ln\mathcal{R}$ functions
when the following likelihood is considered:
\begin{eqnarray}
\mathcal{L}(x, \theta)
&\propto&
\exp(-(x+0.6\theta)^2/(2\cdot1^2))
\nonumber\\
&\times&
\left[\exp(-\theta^2/(2\cdot3^2)\right.
\nonumber\\
&&+\left.0.5\exp(-(x-6)^2/(2\cdot 0.5^2)\right]
\,.
\label{eq:example_llh}
\end{eqnarray}
The dependence of the likelihood on $x$ and $\theta$ is shown in the upper panel of Fig.~\ref{fig:R_vs_L_ex}.
In such case, the $\mathcal{R}$ function takes into account
the existence of a second peak in the posterior.
The choice of the function and the coefficients in Eq.~\eqref{eq:example_llh}
is appropriate to show that, while cutting at 1
(corresponding to the $1\sigma$ limit, in a frequentist sense, for the likelihood ratio test)
the likelihood ratio and the $\mathcal{R}$ methods give the same results,
the cut at 4 (corresponding to a $2\sigma$ significance for the likelihood ratio test) leads to different results,
because the likelihood ratio takes into account only the likelihood values at the best-fit,
while the $\mathcal{R}$ method is also affected by the second peak of the posterior.
For the same reason, the local minimum of $-2\ln\mathcal{R}$ at $x=6$ appears.

Another advantage is computational.
In cosmological analyses, it is typically difficult to study the maximum of the likelihood,
because of the number of dimensions, the numerical noise
and the computational cost of the likelihood.
An example showing the technical difficulties in such kind of analyses
can be found in \cite{Planck:2013nga}.
Similar difficulties can emerge in different analyses.
Even if the best-fit point is not known with sufficient precision, however,
the $\mathcal{R}$ function allows to obtain a prior-independent bound
with a Markov Chain Monte Carlo or similar method.

\section{A simple example with Planck 2018 chains}
\begin{figure}[t]
\centering
\includegraphics[width=\columnwidth]{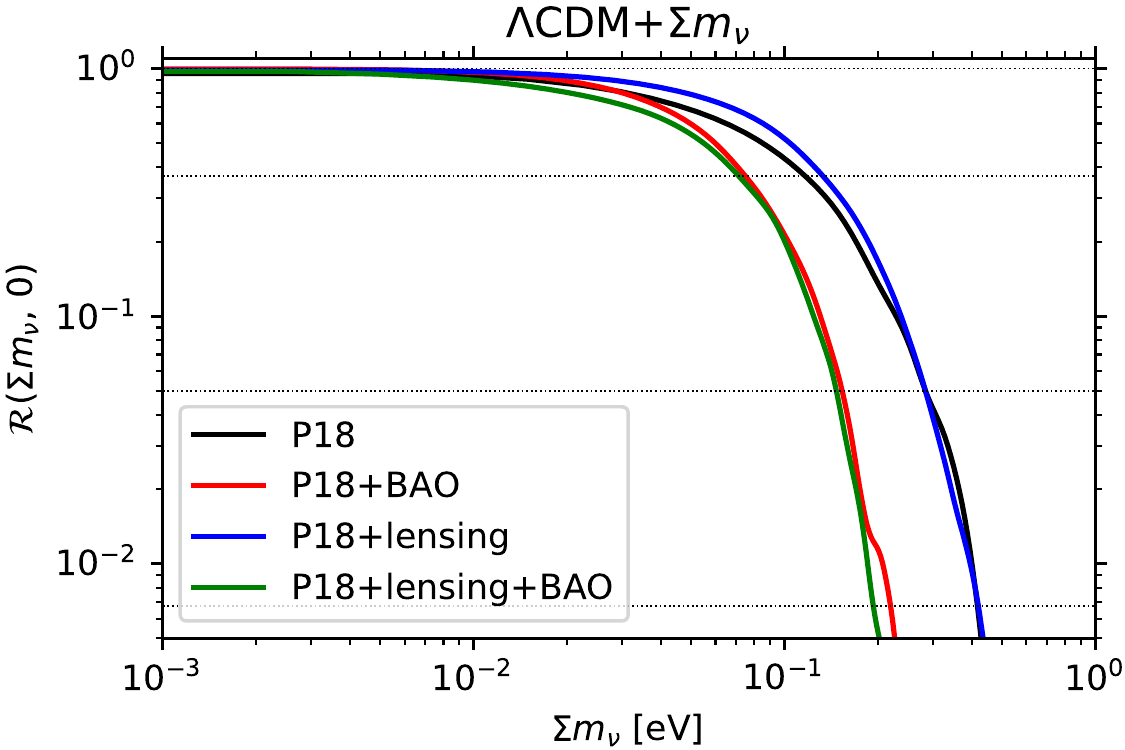}
\caption{\label{fig:R_mnu_planck18}
The $\mathcal{R}(\mnu,\,0)$ function in Eq.~\eqref{eq:R_definition}
obtained from the Planck 2018 chains \cite{Aghanim:2018eyx}
for different data combinations, considering the \lcdm+\mnu\ model.
The horizontal lines show the levels $\ln\mathcal{R}=0$, $-1$, $-3$, $-5$, respectively.
}
\end{figure}

To demonstrate a simple example with recent cosmological data,
we provide in Fig.~\ref{fig:R_mnu_planck18} the function $\mathcal{R}(\mnu,\,0)$ computed in few cases,
obtained from the publicly available Planck 2018 (P18) chains~%
\footnote{The chains are available through the Planck Legacy Archive,
\url{http://pla.esac.esa.int/pla/}.
Note that a simple post-processing of the available chains is sufficient
to produce Fig.~\ref{fig:R_mnu_planck18}.
}
with four different data sets and considering the \lcdm+\mnu\ model.
The datasets include the full CMB temperature and polarization data \cite{Aghanim:2019ame} plus
the lensing measurements \cite{Aghanim:2018oex} by Planck 2018,
and BAO information from
the \texttt{SDSS BOSS} DR12 \cite{Alam:2016hwk,Beutler:2016ixs,Ross:2016gvb,Vargas-Magana:2016imr}
the \texttt{6DF}~\cite{Beutler:2011hx} and
the \texttt{SDSS DR7 MGS}~\cite{Ross:2014qpa} surveys.

The calculation of $\mathcal{R}$ is easy in this case.
The Planck collaboration considered a flat prior on \mnu~%
\footnote{
Note that, although the considered prior is linear,
the calculations through the \texttt{CAMB} code enforce a non-trivial distortion of the prior,
which comes from the numerical requirements of the code.
These come from the fact that some combinations of parameter values may create numerical instabilities,
divergences or simply unphysical values for some cosmological quantities.
These problematic points are therefore excluded by the cosmological calculation ``a priori'',
in the sense that the even if they are formally included in the prior,
their likelihood cannot be computed at the practical level.
In the region below 1~eV, however, the prior on \mnu\ is substantially unchanged by this fact.
},
so we simply have to obtain the posterior $p(\mnu|d,\lcdme+\mnu)$ by standard means
and use it to compute $\mathcal{R}$ according to Eq.~\eqref{eq:R_definition}~%
\footnote{
Note that this is practically what is already shown by most authors in cosmology,
since the results for 1-dimensional marginalized posteriors are often presented in plots
where $p_{\pi}(x|d)/p^{\rm max}_{\pi}$ is shown,
being $\pi(x)$ a linear prior on the quantity $x$,
therefore not affecting the conversion between posterior and $\mathcal{R}$
according to Eq.~\eqref{eq:R_definition}.
Apart for the normalization constant, $p_{\pi}(x|d)/p^{\rm max}_{\pi}$ may be intended
as an unnormalized posterior probability,
which can be employed for bounds calculations as if a linear prior on $x$ was considered,
or as a shape distortion function, therefore not suitable to compute limits unless some prior is assumed first.
}.
Since the lower limit adopted by Planck is $\mnu=0$, we can compute $\mathcal{R}$ for any positive value of \mnu, as far as we do not exceed the upper bound of the prior.
To better show the results,
we consider a logarithmic scale and an appropriate parameter range
for the plot in Fig.~\ref{fig:R_mnu_planck18}.

From the figure, we can notice that the data are completely insensitive
to the value of \mnu\ when it falls below $\simeq0.01$~eV:
in this region, there will be no change between prior and posterior distributions,
and $\mathcal{R}\rightarrow1$ as expected.
On the other hand, $\mnu\gtrsim0.4$~eV will be disfavored by data,
for all the data combinations shown here,
as $\mathcal{R}\rightarrow0$.
As is also expected,
the exact shape of $\mathcal{R}$ between 0.01 and 0.4~eV
depends on the inclusion of the BAO constraints and
only slightly on the lensing dataset.
Regardless of considering a cut at $\mathcal{R}=e^{-3}$ or $\mathcal{R}=e^{-5}$, indeed,
the value of the sensitivity bound only depends on the inclusion of the BAO data.
A comparison of the CMB dataset without (P18) or with (P18+BAO) the BAO constraints, therefore,
can be summarized by two numbers, considering for example $s=-5$:
\begin{eqnarray}
{\mnu}_{,-5} &=& 0.4~\mbox{eV} \qquad \mbox{(P18),}\\
{\mnu}_{,-5} &=& 0.2~\mbox{eV} \qquad \mbox{(P18+BAO).}
\end{eqnarray}

\section{The case of multiple models}
In the previous sections we discussed the case when dealing with only one model,
which was already known in the literature.
The situation is slightly different when more models are considered,
for example if one wants
to study and take into account several extensions of the same minimal scenario,
as in Ref.~\cite{Gariazzo:2018meg}.
It is not difficult to rewrite the definition of $\mathcal{R}$ to deal with multiple models,
if we assume that the prior for the parameter $x$ under consideration is the same in all of them,
i.e.\ that $\pi(x)\equiv\pi(x|\mcm{i})$ does not depend on \mcm{i}.

Let us now recall the method proposed in \cite{Gariazzo:2018meg}.
The model-marginalized posterior distribution of the parameter $x$ is obtained as
\begin{equation}\label{eq:model_marg_posterior}
p(x|d)
=
\sum_i
p(x|\mcm{i},d)\,
p(\mcm{i}|d)\,,
\end{equation}
where
$p(\mcm{i}|d)$ is the posterior probability of the model \mcm{i},
which can be computed using \cite{Handley:2015aa}
\begin{equation}\label{eq:modelposterior}
p(\mcm{i}|d)
=
\frac{Z_i\pi(\mcm{i})}{\sum_j Z_j\pi(\mcm{j})}
\,.
\end{equation}
In both cases the sum runs over all the studied models.
Coming back to Eq.~\eqref{eq:model_marg_posterior} and using Eqs.~\eqref{eq:marginalizedposterior_explicit}~%
\footnote{
The marginalization over the parameters $\psi$ is not necessarily the same in all the models.
As we are not assuming anything on $\psi$,
they can be not the shared ones among the various models and can vary in number.
In any case, the marginalization works inside each model independently,
using for each \mcm{i} the appropriate parameter space and priors:
the differences remain hidden in the definition of $Z_i^x$.
}
and \eqref{eq:modelposterior},
we obtain the fully (prior- and model-) marginalized
posterior probability of $x$:
\begin{equation}\label{eq:fully_marg_post_x}
p(x|d)
=
\frac{
\sum_i
\pi(x|\mcm{i})
Z_i^x
\pi(\mcm{i})}
{\sum_j
Z_j
\pi(\mcm{j})}\,.
\end{equation}
Remembering that we assumed $\pi(x)\equiv\pi(x|\mcm{i})$ to be independent of \mcm{i},
the ratio between prior and marginalized posterior probabilities
for the parameter $x$ is:
\begin{equation}\label{eq:fully_marg_post_over_pri}
\frac{p(x|d)}{\pi(x)}
=
\frac{
\sum_i
Z_i^x
\pi(\mcm{i})
}
{\sum_j
Z_j
\pi(\mcm{j})}\,.
\end{equation}
If we use this result to define $\mathcal{R}$ again as in equation~\eqref{eq:R_definition},
we have:
\begin{equation}\label{eq:R_definition_mm}
\mathcal{R}(x,x_0|d)
\equiv
\frac
{\sum_iZ_i^{x}\pi(\mcm{i})}
{\sum_jZ_j^{x_0}\pi(\mcm{j})}
=
\frac{p(x|d)/\pi(x)}{p(x_0|d)/\pi(x_0)}
\,,
\end{equation}
which has exactly the same meaning as before,
apart for the fact that in this case $\mathcal{R}$ has been marginalized over several models.

From the computational point of view, in the model-marginalized case obtaining $\mathcal{R}$
is as simple as when only one model is considered.
One just has to select a prior $\pi(x)$ and a sufficiently broad range,
obtain the marginalized posterior probability as in \cite{Gariazzo:2018meg},
then divide it by the considered prior and normalize appropriately.

\begin{figure}[t]
\centering
\includegraphics[width=\columnwidth]{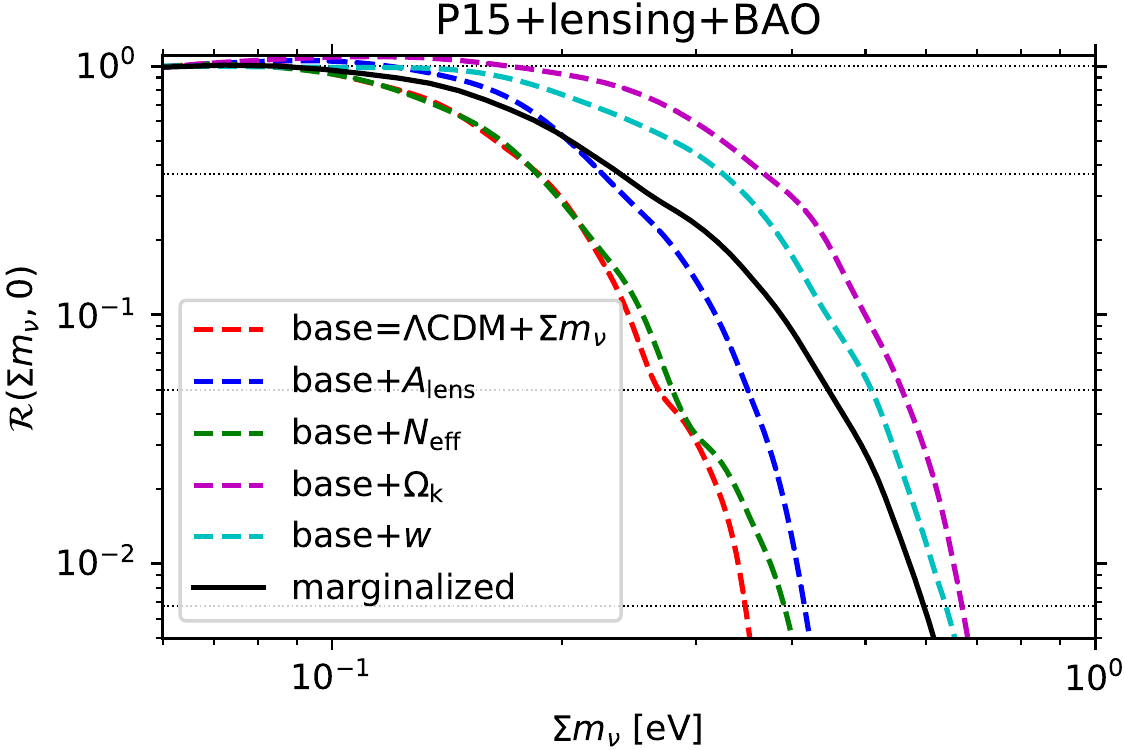}
\caption{\label{fig:R_mnu_mm}
The $\mathcal{R}(\mnu,\,0)$ function in Eq.~\eqref{eq:R_definition}
obtained considering different models (dashed)
together with the model-marginalized one from Eq.~\eqref{eq:R_definition_mm} (solid),
using the full dataset adopted in Ref.~\cite{Gariazzo:2018meg} (see text for details).
The horizontal lines show the levels $\ln\mathcal{R}=0$, $-1$, $-3$, $-5$, respectively.
}
\end{figure}

As an example, we provide in Fig.~\ref{fig:R_mnu_mm} the $\mathcal{R}$ function
obtained from the vary same posteriors studied in Ref.~\cite{Gariazzo:2018meg}.
Such cases are computed considering the full Planck 2015 (P15) CMB data \cite{Adam:2015rua,Ade:2015xua},
together with the lensing likelihood \cite{Ade:2015zua}
and the BAO observations by
from the \texttt{SDSS BOSS} DR11 \cite{Anderson:2013zyy},
the \texttt{6DF}~\cite{Beutler:2011hx} and
the \texttt{SDSS DR7 MGS}~\cite{Ross:2014qpa} surveys.
The considered models are the same extensions of the $\Lambda$CDM+\mnu\ case
adopted by the Planck collaboration
for the 2015 public release,
but with a prior $\mnu>60$~meV.
Also in this case we can see how the $\mathcal{R}$ function is very close to one below 0.1~eV
and always goes to zero above $\sim0.7$~eV.
In the middle, the various models (dashed lines) have different constraining powers,
whose weighted average is represented by the solid line.
The model-marginalized, prior-independent result corresponds to
\begin{equation}
{\mnu}_{,-5} = 0.6~\mbox{eV} \qquad \mbox{(P15+lensing+BAO).}\\ 
\end{equation}

\section{Discussion and conclusions}
In this letter we discussed a possible way to show
prior-independent results in the context of Bayesian analysis,
generalising a previously known method \cite{Astone:1999wp,DAgostini:2000edp,DAgostini:2003}
to deal with multiple models, extending also the work presented in Ref.~\cite{Gariazzo:2018meg}.
The method uses Bayesian model comparison techniques to compare the constraining power
of the data at different values of the considered parameter, and is particularly useful
when open likelihoods are involved in the analysis.
While the method can be similar to a likelihood ratio test,
it does not only take into account the information contained in the best-fit point,
i.e.\ the maximum of the likelihood,
but also the information of the full posterior,
so that in case of multivariate posterior distributions, more conservative limits are obtained.
Furthermore, the discussed method can be much less expensive than the likelihood ratio test
from the computational point of view.

We applied the simple formulas to the case of neutrino mass constraints from cosmology,
discussing the case of several datasets analyzed with one single cosmological model,
and the case where we have only one dataset but multiple models.
In the latter case, Bayesian model comparison also allows
to take into account the constraints from the different models
to obtain a prior-independent and model-marginalized bound.
An extended application of this method is left for a separate work.

\begin{acknowledgements}
The author receives support from the European Union's Horizon 2020 research and innovation programme under the Marie Sk{\l}odowska-Curie individual grant agreement No.\ 796941.
\end{acknowledgements}

\bibliography{main}

\end{document}